\documentclass[preprint,prl,aps,superscriptaddress,showpacs]{revtex4}
\usepackage{graphicx}
\begin{document}
\title
{{\it Ab initio} calculation of intrinsic spin Hall effect in semiconductors}
\author{G. Y. Guo}
\email{gyguo@phys.ntu.edu.tw}
\affiliation{Department of Physics, National Taiwan University,
Taipei 106, Taiwan}
\author{Yugui Yao}
\affiliation{Institute of Physics, Chinese Academy of Sciences, Beijing 100080, China}
\author{Qian Niu}
\affiliation{Department of Physics, University of Texas at Austin, 
Austin TX 78712-1081, USA}

\date{\today}   %submitted to PRL via postal mail
\begin{abstract}
Relativistic band theoretical calculations reveal that  
intrinsic spin Hall conductivity in hole-doped archetypical semiconductors 
Ge, GaAs and AlAs is large $[\sim 100 (\hbar/e)(\Omega cm)^{-1}]$,
showing the possibility of spin Hall effect beyond the four band Luttinger Hamiltonian.
The calculated orbital-angular-momentum (orbital) Hall conductivity is 
one order of magnitude smaller, indicating no cancellation between
the spin and orbital Hall effects in bulk semiconductors.
Furthermore, it is found that the spin Hall effect can be strongly 
manipulated by strains, and that the $ac$ spin Hall 
conductivity in the semiconductors is large in pure as well as doped semiconductors.
\end{abstract}
\maketitle

%\section{Introduction}
Spin-orbit coupling is known to be responsible for a whole range of
interesting phenomena in magnetic materials~\cite{ebe96a}, including magnetic
anisotropy, optical dichroism, and anomalous Hall effect.
Spin generation and transport in paramagnetic materials can also be induced
by an electric field because of spin-orbit coupling even in the absence
of a magnetic field as demonstrated in a number of recent 
experiments~\cite{aws04}. This offers the exciting possibility of pure electric
driven spintronics in semiconductors, where spin-orbit coupling is relatively
strong and which can be more readily integrated with conventional 
electronics~\cite{wol01}. Earlier theories on electric spin generation and 
transport were based on extrinsic effects due to spin-orbit coupling 
in scattering processes~\cite{dya71}. Recently, it has been proposed
that a transverse spin current response to an electric field, known as the intrinsic spin
Hall effect, can also occur in pure crystalline materials due to intrinsic
spin-orbit coupling in the band structure~\cite{mur03,sin04}. 

A large number of theoretical papers have been written addressing various issues 
about the intrinsic spin Hall effect.  In \cite{cul04}, a systematic semi-classical 
theory of spin transport is presented, resolving a discrepancy between the prediction 
of \cite{mur03} and the Kubo formula result.  In \cite{zha04}, an orbital-angular-momentum 
(orbital) Hall current is predicted to exist in response to an electric field 
and is found to cancel exactly the spin Hall current in the spin Hall effect.  
There is also an intensive debate about whether the intrinsic spin Hall effect 
remains valid beyond the ballistic transport regime \cite{ino04}.  
On the other hand, experimental measurements of large spin Hall effects for the Rashba two 
dimensional electron gas and for n-type bulk semiconductors have just been reported~\cite{wun04,kat04}, 
although more work is needed to firmly establish the intrinsic or extrinsic nature of the results. 
It is expected that similar experiments on p-type semiconductors will also appear soon. 

Inspired by the prospect that the intrinsic spin Hall effect may provide 
a useful property for spintronics applications which is designable based 
on material parameters, we have carried out {\it ab initio} calculations 
on this effect and related phenomena in the archetypical p-type semiconductors 
Si, Ge, GaAs and AlAs.  We focus on the p-type because the spin-orbit coupling
is much stronger in the valence bands, where the intrinsic effect
has a better chance of dominating over the extrinsic scattering effects.
Such results are clearly needed to provide a stage 
for systematic and quantitative comparison between theoretical predictions 
and experimental measurements.  In addition, the calculations may also help 
resolving some of the theoretical issues for such semiconductors whose 
discussions are currently all based on the four-band Luttinger Hamiltonian 
for the holes and often with the spherical approximation.  

Our results cover a large range of hole concentration which is beyond the 
validity regime of the four band model. By including all the relevant bands, 
we find a pronounced spin Hall conductivity for all the semiconductors except 
the light element Si and a roughly quadratic dependence of the spin Hall conductivity 
on the spin-orbit gap.  The vanishing of spin Hall conductivity in the limit 
of zero spin-orbit coupling differs qualitatively with the prediction 
of the four band model, which yields a finite spin Hall conductivity even 
when spin-orbit coupled terms in the Luttinger Hamiltonian approach zero~\cite{mur03,wan04}. 
We also calculated the orbital Hall effect, and find it an order of magnitude 
weaker than the spin Hall effect, a result due to the orbital quenching by 
the cubic anisotropy of the crystals.  This is in sharp contrast with the 
case of the Rashba two-dimensional electron system and the spherical four-band model, 
where exact cancellation between the spin and orbital Hall effects occurs 
due to rotational symmetry in such models~\cite{zha04}.  In addition, we 
calculated the effect of strain, which is commonly present in semiconductor 
multilayers and superlattices due to lattice mismatch, and which may be used 
to enhance or reduce the spin Hall effect. Furthmore, this result on strain effect will 
help the experimentalists to distinguish intrinsic from extrinsic aspect of
the spin Hall effect in the future~\cite{kat04}. Finally, we also calculated 
the $ac$ spin Hall conductivity and find it to be large in both pure and 
hole-doped semiconductors. 
%In contrast, the $ac$ orbital Hall conductivity 
%for pure semiconductors is at least two order of magnitude smaller than 
%the $ac$ spin Hall conductivity.

%\section{Theorey and Computational details}
The intrinsic Hall conductivity of a solid can be evaluated by using
the Kubo formula approach~\cite{mar00}.
In this approach, the intrinsic Hall effect comes from the 
static $\omega = 0$ limit of the off-diagonal element of 
the conductivity tensor~\cite{mar00,sin04}:
\begin{eqnarray}
 \sigma_{xy}(\omega) =-\frac{e}{i\omega V_c}
 \sum_{\bf k}\sum_{n\neq n'} (f_{{\bf k}n}-f_{{\bf k}n'})\nonumber \\
\frac{<{\bf k}n|j_x|{\bf k}n'><{\bf k}n'|v_y|{\bf k}n>}
 {\epsilon_{{\bf k}n}-\epsilon_{{\bf k}n'}+\hbar\omega+i\eta}
\end{eqnarray}
where $V_c$ is the unit cell volume, $\hbar\omega$ is the photon energy,
and $|{\bf k}n>$ is the $n$th Bloch state with
crystal momentum ${\bf k}$. Since all the intrinsic Hall effects
are caused by spin-orbit coupling, first-principles
calculations must be based on a relativistic band theory. In this
case, the current
operators ${\bf j}$ are $-e{\bf v} $, 
$\frac{\hbar}{4}\{\beta\Sigma_z, {\bf v}\}$,
and $\frac{\hbar}{2}\{\beta L_z, {\bf v} \}$ for the anomalous, spin, and
orbital Hall effects, respectively. 
Here $\beta, \Sigma$ are the well-known $4\times 4$ Dirac 
matrices~\cite{ber71}, and ${\bf v}$ is the velocity operator projected 
onto states above the electron-positron mass gap \cite{guo95}.
Setting $\eta$ to zero and using $Im[1/(x-i\eta)] = \pi\delta(x)$,
one obtains the imaginary part of the off-diagonal element
\begin{eqnarray}
 \sigma''_{xy}(\omega) =\frac{\pi e}{ \omega V_c}
 \sum_{\bf k} \sum_{n\neq n'} (f_{{\bf k}n}-f_{{\bf k}n'})\\
 Im[<{\bf k}n|j_x|{\bf k}n'><{\bf k}n'|v_y|{\bf k}n>]
 \delta(\epsilon_{n' n}-\hbar\omega) \nonumber
\end{eqnarray}
where $\epsilon_{n' n} = \epsilon_{{\bf k}n'}- \epsilon_{{\bf k}n}$.
As in previous magneto-optical calculations~\cite{guo95}, 
we first calculate the imaginary part of the $\sigma_{xy}$
and then obtain the real part from the imaginary
part by a Kramers-Kronig transformation
\begin{equation}
 \sigma'_{xy}(\omega) = \frac{2}{\pi}{\bf P} \int_0^{\infty}d\omega'
 \frac{\omega'\sigma''_{xy}(\omega')}{\omega'^2-\omega^2}
\end{equation}
where ${\bf P}$ denotes the principal value of the integral.
We notice that the anomalous Hall conductivity $\sigma_{xy}(0)$
of bcc Fe calculated this way~\cite{guo95,yao04} is in good quantitative
agreement with that calculated directly by a generalized Boltzmann
equation approach that accounts for the Berry phase correction to the
group velocity~\cite{yao04}. 
%Furthermore, we have also calculated
%the spin Hall conductivity for Ge by using direct summation, i.e.,
%using Eq. (1) in its static $\omega$ = 0 limit, and the results
%agree well (within 5\%) with that obtained through Kramers-Kronig
%transformation.

\begin{table}[h]
    \caption{\label{lattice} Experimental lattice constant $a$
(see ~\cite{che87} and references therein),
   average atomic sphere radius $R_{ws}$ and band gap $E_g$
(see ~\cite{joh98} and references therein)
of the semiconductors
studied. The calculated band gaps $E_g^{the}$ and
spin-orbit splitting $\Delta_{so}$ of the top valence bands
at $\Gamma$ are also listed.}
 \begin{center}
        \begin{tabular}{cccccc}
\hline \hline
     & $a$ (\AA)& $R_{ws} (a.u.) $ & $E_g$ (eV) & $E_g^{the}$ (eV)&
 $\Delta_{so}$ (meV) \\
\hline
Si   & 5.431   & 2.526 & 1.17 & 0.81 & 47\\
Ge   & 5.650   & 2.632 & 0.74 & 0.28 & 278\\
AlAs & 5.620   & 2.615 & 2.23 & 1.52 & 301\\
GaAs & 5.654   & 2.632 & 1.52 & 0.76& 336\\
%GaSb & 6.118   & 2.846 & 0.81 & 0.10& 591\\
%InAs & 6.036   & 2.808 & 0.42 & 0.15& 318\\
%InSb & 6.478   & 3.014 & ? & 0.03 & 526\\
\hline \hline
        \end{tabular}
    \end{center}
\end{table}

\begin{figure}[h]
\includegraphics[width=8cm]{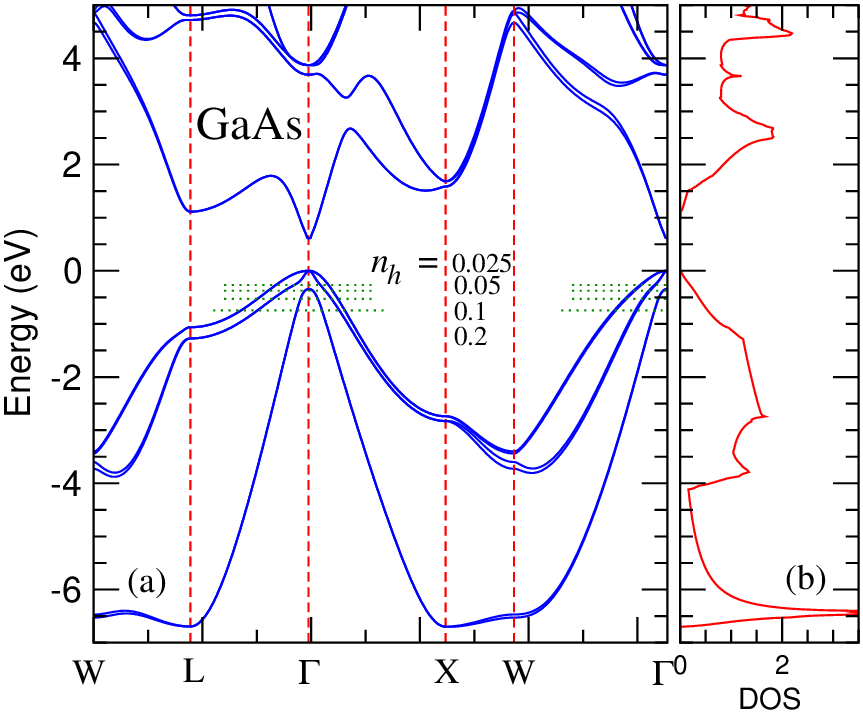}
\caption{\label{bs} (color online) (a) Relativistic
 band structure of GaAs: (a) Energy bands and (b)
 density of states (DOS) (states/eV/cell). 
 The zero energy is at the top of 
 the valence bands. The dotted lines just below the top of
 the valence bands denote the Fermi levels for the
 hole concentration $n_h$ = 0.025, 0.05, 0.1, and 0.2
 e/cell, respectively. }
%    \speci al{psfile=pdosb.fig
%        hoffset=0 voffset=-20 hscale=32 vscale=35 angle=-90}
\end{figure}

The relativistic band structure of the semiconductors is
calculated using a fully relativistic extension~\cite{ebe88}
of the well established all-electron linear muffin-tin orbital
(LMTO) method~\cite{and75}. The calculations are based
on density functional theory with generalized gradient approximation
(GGA)~\cite{per96}. In the fcc diamond or zincblend structures,
two atoms sit at (0,0,0) and (1/4,1/4,1/4), respectively.
In the present calculations, two "empty" atomic spheres 
are introduced at the vacant sites (1/2,1/2,1/2) and (3/4,3/4,3/4),
respectively, to make the structures more close-packed. The
four "atoms" in the unit cell are assumed to have an equal
atomic sphere radius. The experimental
lattice constants for Si, Ge, GaAs and AlAs used are listed in
Table I together with the corresponding atomic sphere
radii. The basis functions used for all the "atoms" are $s$,
$p$, and $d$ muffin-tin orbitals~\cite{and75}.
In the self-consistent electronic structure calculations, 
eighty-nine $k$-points in the fcc irreducible wedge (IW)
of the Brillouin zone (BZ) were used in the tetrahedron
BZ integration. The calculated band gaps $E_g$ and spin-orbit
splitting $\Delta_{so}$ of the top valence bands at $\Gamma$ for all the
semiconductors are listed in Table I.
As an example, the calculated band structure of GaAs
is displayed in Fig. 1.

In the Hall conductivity calculations, 
a much finer $k$-point mesh is needed. Moreover, a larger IW 
(three times the fcc IW for Si and Ge, and six times the
fcc IW for AlAs and GaAs which have no spatial inversion
symmetry) is necessary for the Hall conductivity calculations.
The number of $k$-points in the IW used is 49395 for Si and Ge 
and 98790 for GaAs and AlAs. These numbers are obtained by
dividing the $\Gamma X$ line into 56 intervals (see Fig. 1).
Test calculations using 139083 $k$-points (80 divisions of the $\Gamma X$ line) 
for Ge indicate that the calculated spin Hall
conductivity converge to within 2 \%. The imaginary part
of the Hall conductivity is calculated up to 50 eV to
ensure that the real part of the conductivity obtained through
the Kramers-Kronig transformation converges as well.
As is well known, all the calculated band gaps are smaller
than the corresponding experimental values though
the calculated spin-orbit splittings $\Delta_{so}$ 
agree quite well with the experimental ones as well
as the full-potential calculations (see Table I and ~\cite{car04}). 
To remedy this defect of the GGA, the so-called scissor operator
is applied in the conductivity calculations, i.e., all the
conduction bands are shifted upwards such that the theoretical
band gap is the same as the experimental one.  

\begin{figure}[h]
\includegraphics[width=8cm]{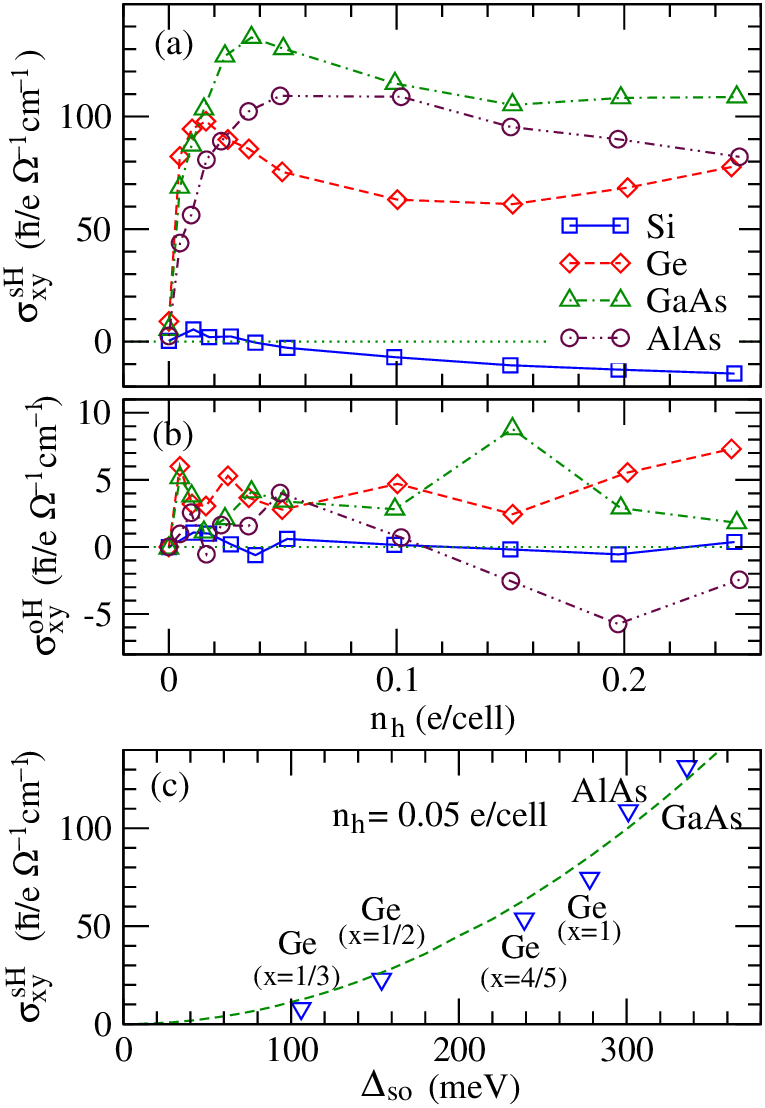}
\caption{\label{sh} (color online) Calculated (a) spin 
 ($\sigma_{xy}^{sH}$) and (b)
 orbital ($\sigma_{xy}^{oH}$) Hall
 conductivity as a function of hole concentration ($n_h$) for
 Si, Ge, GaAs and AlAs. (c) Calculated spin Hall conductivity
 for $n_h = 0.05$ e/cell vs. the top valence band spin-orbit splitting
 ($\Delta_{so}$). The dashed line denotes the parabola fitted to
 the calculated data. The $x$ is the spin-orbit coupling 
 scaling factor~\cite{ebe96}: $x = 1$ means full spin-orbit coupling
 and $x = 0$ means no spin-orbit coupling. $n_h = 0.1$ e/cell is 
 equivalent to $n_h$ of 2.5$\times 10^{21}$, 2.2$\times 10^{21}$, 
 2.2$\times 10^{21}$ and 2.3$\times 10^{21}$ cm$^{-3}$ for Si, Ge,
 GaAs and AlAs, respectively.}
\end{figure}

%\section{Results and discussion}
Fig. 2 shows the calculated spin ($\sigma_{xy}^{sH}$)
and orbital ($\sigma_{xy}^{oH}$) Hall conductivities
as a function of hole concentration. 
The spin Hall conductivity in Ge, GaAs and AlAs
increases sharply as 
the hole concentration $n_h$ increases at small  $n_h$ 
($n_h \leq$  0.02 e/cell) (Fig. 2(a)), being consistent with the recent analytical 
Luttinger model predictions of $n_h^{1/3}$~\cite{cul04,mur04}.
However, it becomes more or less
flat when $n_h$ is further increased (Fig. 2(a)).
This is due to the fact that the spin Hall effect results 
predominantly from the spin-orbit split heavy-hole (HH) and light-hole (LH) 
pockets centred at $\Gamma$ (Fig. 1)~\cite{mur03}.
Fig. 1 shows that when $n_h \geq$  0.05 e/cell, the Fermi
level is already below the top of the lower spin-orbit
split valence band. However, there is no substantial increase
in the spin Hall conductivity as $n_h$ goes through 0.05 e/cell,
indicating that the lower spin-orbit
split valence band has no significant contribution to
the spin Hall conductivity.
Remarkably, the spin conductivity for Ge, GaAs and AlAs is very pronounced,
being about 100 $\hbar /(e \Omega cm)$ (Fig. 2). These values are
comparable to the charge conductivity for lightly hole-doped 
Ge, GaAs and AlAs, in agreement with
the prediction of the large intrinsic spin Hall effect in 
hole-doped semiconductors based on the Luttinger Hamiltonian \cite{mur04,cul04}. 
However, our results differ from the predictions of the four band model 
in at least one important aspect. Fig. 2(c) shows that the
calculated spin Hall conductivity at, e.g., $n_h = 0.05$ e/cell, 
approaches zero nearly quadratically as the spin-orbit splitting
$\Delta_{so}$ and hence the spin-orbit
coupling strength goes to zero. 
In contrast, the four band model yields a finite spin Hall conductivity even 
as the spin-orbit coupled terms in the Luttinger Hamiltonian approach zero.  

\begin{figure}[h]
\includegraphics[width=8cm]{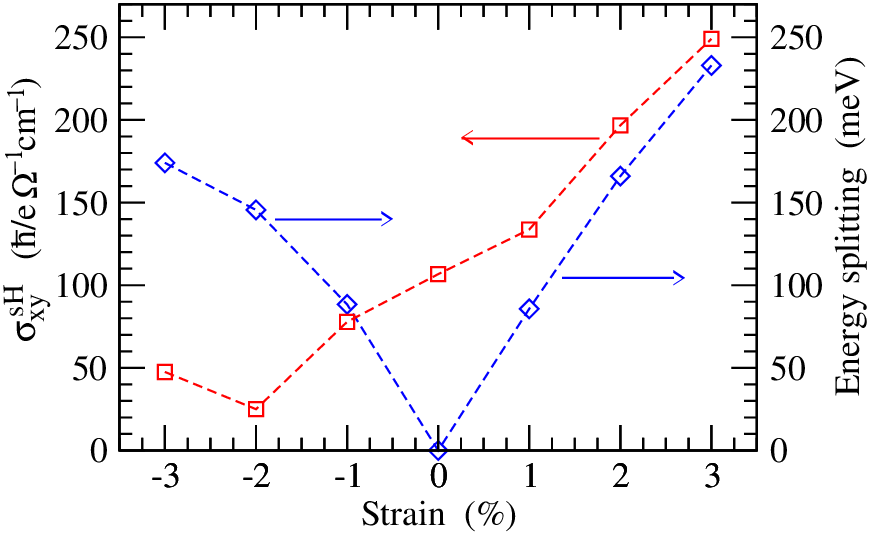}\\
\caption{\label{sigma} (color online) Calculated spin Hall
  conductivity ($\sigma_{xy}^{sH}$) (squares) and
  energy splitting (diamonds) of heavy-hole and light-hole bands at $\Gamma$
  of GaAs as a function of uniaxial strain at
  the hole concentration $n_h = 0.1 $ (e/cell).}
\end{figure}

The calculated orbital Hall conductivity is at least ten times 
smaller than the spin Hall conductivity (Fig. 2), demonstrating
that the orbital Hall effect in bulk semiconductors is strongly
quenched by the cubic crystal fields and covalent bonding. 
%Furthermore, except for Si,
Furthermore, the orbital and spin Hall conductivities generally have the same sign when the hole
concentration is small, i.e., $n_h \leq 0.1$ e/cell (Fig. 2).
Therefore, the exact cancellation of the intrinsic spin Hall effect
by the intrinsic orbital Hall effect in the two-dimensional electron gas with
the Rashba spin-orbit term predicted in ~\cite{zha04}
would not occur in bulk semiconductors.
%We have also calculated the Hall
%conductivities for the semiconductors at
%several doped-electron concentrations and 
%the doped electron concentration $n_e$ = 0.05, 0.1 e/cell
%find both the spin and orbital Hall conductivities to be small.
%This finding is consistent with the first observed spin Hall
%effect in bulk $n$-type semiconductors being extrinsic~\cite{kat04}.
%negligibly small, indicating that the electron pockets in the low lying conduction
%bands do not contribute to the spin Hall and oH conductivities.
%This may be expected because in the conduction band region,
%there is no quadruply degenerate singular
%point like the HH and LH bands at $\Gamma$ (Fig. 1).
 
In device applications, semiconductors are typically grown 
epitaxially on substrates with similar lattice constants, resulting
in semiconductor multilayers and superlattices. The semiconductor
layers are thus generally strained due to small lattice mismatches.
The lattice mismatch strain may have significant effects on the
electronic structure of the semiconductors, and hence also on the spin Hall
effect~\cite{ber04}. In particular, the degeneracy of the HH and LH
bands at $\Gamma$ would be lifted. We have calculated 
the band structure and Hall conductivity of GaAs under [001]
uniaxial elastic strains. The calculated spin Hall conductivity
as well as the energy splitting $\Delta$ of the HH and LH bands
at  $\Gamma$ are displayed as a function of strain in Fig. 3.  
Here the strain $\varepsilon$ is defined as
$a = a_0(1+\varepsilon)$ where $a$ and $a_0$ are the strained
and unstrained lattice constants perpendicular to [001], respectively.
As expected, the $\Delta$ is roughly proportional to 
the strain size. Remarkably, the spin Hall conductivity
varies almost linearly with the strain, increasing with tensile change 
and decreasing with compression in the lateral directions (Fig. 3). 
This linear and sensitive dependence on the strain may be used to 
tune the spin Hall effect.

\begin{figure}[h]
\includegraphics[width=7cm]{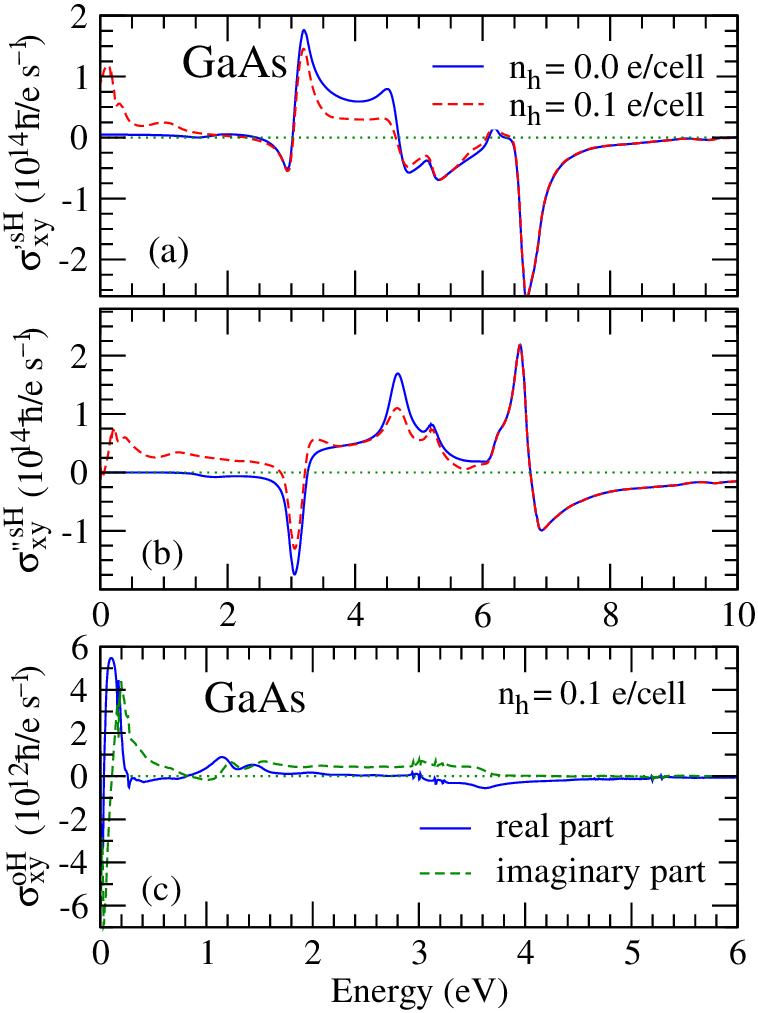}\\
\caption{\label{sigma} (color online) Calculated spin ($\sigma_{xy}^{sH}$) and
  orbital ($\sigma_{xy}^{oH}$) Hall
  conductivity of GaAs as a function of frequency at the 
  hole concentrations $n_h = 0.0, 0.1 $ (e/cell), respectively.}
\end{figure}

%Let us now consider the $ac$ Hall conductivities. 
The shape of the calculated
$ac$ spin Hall conductivity spectra for all the semiconductors studied
here look rather similar, though the peak energy position and peak magnitudes
may differ. Thus, we only display in Fig. 4 the
$ac$ Hall conductivities for GaAs as a representative.
Interestingly, both the real and imaginary parts of the spin Hall
conductivity in both pure and hole-doped semiconductors are large 
(Fig. 4a-b). This suggests that one could generate 
$ac$ spin current in semiconductors without using magnetic field or magnetic
materials. In contrast, the orbital Hall conductivity in the
pure semiconductors is zero (i.e., within
numerical noise) whilst it is two order of magnitude smaller than the
spin Hall conductivity in the hole-doped semiconductors (Fig. 4).

%\section{Conclusions}
%To summarize, we have investigated intrinsic spin and orbital Hall 
%effects in archetypical semiconductors Si, Ge, GaAs and AlAs by first-principles relativistic band 
%theoretical calculations. The calculated intrinsic spin Hall
%conductivity for Ge, GaAs and AlAs is large, thereby showing the existence of the
%intrinsic spin Hall effect in the hole-doped semiconductors beyond the
%Luttinger Hamiltonian. The calculated orbital Hall conductivity 
%is one order of magnitude
%smaller than the spin Hall conductivity, ruling out the suggestion that
%the spin Hall effect would be fully compensated by the orbital Hall effect.
%Furthermore, we find that the lattice mismatch strains in semiconductor
%multilayers can be used to control the spin Hall effect and that
%the $ac$ spin Hall conductivity in the semiconductors is
%also large.

The authors thank Ming-Che Chang for stimulating discussions.
They thank NCTS/TPE and National Science Council of ROC, NSFC 
(No. 10404035) of PRC, US Department of Energy (DE-FG03-02ER45958) 
and the Welch Foundation for financial supports, 
and thank NCHC of ROC for providing CPU time. \\

\end{document}